\documentclass{aastex63} 

\usepackage{epsf}
\usepackage{amsmath}                
\usepackage{amsfonts}               
\usepackage{amssymb}                
\usepackage{epsfig}                 
\usepackage{float}
\usepackage{color}
\usepackage{tabularx}
\usepackage{comment}
\usepackage{makecell}

\def \MeV{~\rm{MeV}}

\def \eV{~\rm{eV}}

\def \cm{~\rm{cm}}
\def \s{~\rm{s}}
\def \km{~\rm{km}}

\def \K{~\rm{K}}

\def \g{~\rm{g}}
\def \G{~\rm{G}}

\def \yr{~\rm{yr}}

\def \keV{~\rm{keV}}

\begin{document}


\title{Mergers of compact objects with cores of massive stars: evolutionary pathways, r-process nucleosynthesis and multi-messenger signatures}

\author[0000-0002-2215-1841]{Aldana Grichener}
\affiliation{Steward Observatory, University of Arizona, 933 North Cherry Avenue, Tucson, AZ 85721, USA; agrichener@arizona.edu}

\begin{abstract}

The study of massive binary systems has steadily progressed over the past decades, with increasing focus on their evolution, interactions and mergers, driven by improvements in computational modelling and observational techniques. In particular, when a binary system involves a massive giant and a neutron star (NS) or a black hole (BH) that go through common envelope evolution (CEE), it might result in the merger of the compact object with the core of its giant companion, giving rise to various high energy astrophysical phenomena. We review the different evolutionary channels that lead to compact object-core mergers, key physical processes with emphasis on the role of accretion physics, feasibility of r-process nucleosynthesis, expected observable electromagnetic, neutrino and gravitational-wave (GW) signatures, as well as potential correlation with detected core collapse supernovae (CCSNe), luminous fast blue optical transients (LFBOTs) and low luminosity long gamma-ray bursts (LGRBs). After presenting our current understanding of these mergers, we conclude discussing prospects for future advancements.   

\end{abstract}


\section{INTRODUCTION}
\label{sec:intro}

Compact objects are frequently found in binary systems. When a neutron star (NS) or a black hole (BH) is close enough to a massive companion, the immense swelling of the companion as it evolves might result in the engulfment of the compact object that starts spiralling inside its envelope due to dynamical friction. This evolutionary phase, which is commonly known as common envelope evolution (CEE; e.g.,  \citealt{Paczynski1976CEE}; \citealt{TaamBodenheimer1978CE}; \citealt{IbenLivio1993CEE}; \citealt{Izzardetal2012CEE}; \citealt{Ivanovaetal2013CEE}; \citealt{Ivanovaetal2020CEEbook}; \citealt{RopkeDeMarco2023CEEreview}), could result either in the ejection of the envelope and survival of the compact object-core binary (e.g., \citealt{VignaGomezetal2018DNSs}), or their merger within the envelope gas (e.g., \citealt{Grichener2023popSynth}). The vast accretion by the compact object as it merges with the core of the giant could lead to the formation of an accretion disk (e.g., \citealt{Papishetal2015Rprocess}; \citealt{GrichenerSoker2019rprocess};  \citealt{Eversonetal2024TZOs}). In cases where jets launched by this accretion disk play a key role in the merger, the resultant transient is termed \textit{common envelope jets supernova} (CEJSN; \citealt{SokerGilkisiPTF14hls}). 

Mergers of NSs/BHs with the cores of their giant companions have been attracting interest as potential r-process nucleosynthesis sites (e.g., \citealt{Papishetal2015Rprocess}; \citealt{GrichenerSoker2019rprocess}; \citealt{Gricheneretal2022GCE}; \citealt{Grichener2023popSynth}; \citealt{JinSoker2024Rprocess}) and potential progenitors of long gamma-ray bursts (LGRBs; e.g., \citealt{FryerWoosley1998HeStarBHmerger}; \citealt{ZhangFryer2001HeStarBHmerger}; \citealt{Thoneetal2011GRB101225A}; \citealt{RuedaRuffini2012}; \citealt{Fryeretal2014GRBs}), merger driven core collapse supernovae (CCSNe; e.g., \citealt{Fryeretal1996MergerDrivenExplosions}; \citealt{BarkovKomissarov2011hypernova}; \citealt{Chevalier2012IIn}; \citealt{Schroderetal2020mergers}; \citealt{SokerGilkisiPTF14hls}; \citealt{Gofmanetal2019iPTF14hls}; \citealt{Dongetal2021mergerTriggeredCCSN}) and luminous fast blue optical transients (LFBOTs; e.g., \citealt{Sokeretal2019FBOTs}; \citealt{Metzger2022FBOTs}; \citealt{Soker2022FBOTs}; \citealt{CohenSoker2023FBOTs}).  Due to their complexity and the wide variety of physical processes they encompass, many studies perform semi-analytical and one-dimensional simulations to explore these mergers (e.g., \citealt{Sokeretal2019FBOTs}; \citealt{GrichenerSoker2019rprocess}; \citealt{GrichenerSoker2021neutrinos}; \citealt{CohenSoker2023FBOTs}), while others simulate selected aspects in three dimensions (e.g., \citealt{Schroderetal2020mergers}; \citealt{Schreieretal2021CEJSN}; \citealt{Hilleletal2022CEJSN}; \citealt{Hilleletal2022NJF3D}; \citealt{Eversonetal2024TZOs}). However, to date, computational barriers do not allow for fully self-consistent simulations of these transients. The aim of this review is to present our current understanding of NS/BH-core mergers and explore the role they play in high energy astrophysics, while highlighting the limitations of the currently performed studies, and presenting future directions that are essential for addressing the open questions that remain in this field.   

This review is organized as follows. We present the possible evolutionary paths towards NS/BH-core mergers (section \ref{sec:channels}) and discuss key physical processes relevant to these mergers and their effect on the accretion rate by the compact object (section \ref{sec:physics}). We then present studies of r-process nucleosynthesis in NS-core merger (section \ref{sec:rprocess}). In section \ref{sec:observations} we elaborate on the expected observational properties, including electromagnetic signatures and correlation with detected transients (section \ref{subsec:EMsignals}), neutrinos (section \ref{subsec:neutrinos}) and gravitational-waves (GWs; section \ref{subsec:GWs}). We give future prospects in section \ref{sec:future} and summarize in section \ref{sec:Summary}.    

\section{Evolutionary channels}
\label{sec:channels}


\begin{figure*}
\begin{center}
\includegraphics[width=0.9\textwidth]{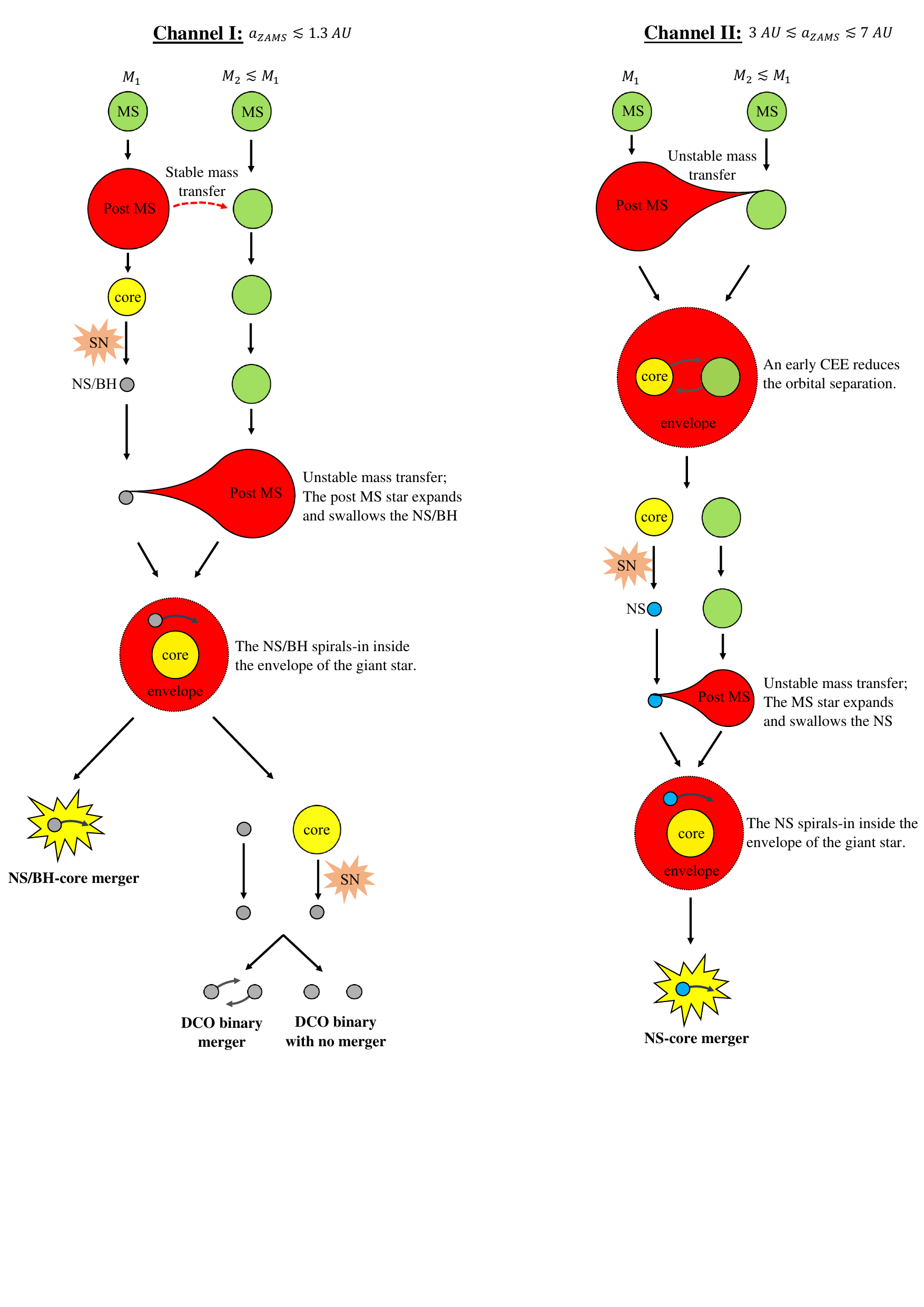}
\vspace*{-3.8cm}
\caption{Main evolutionary routes towards NS/BH - core mergers from \cite{Grichener2023popSynth}. In both channels the primary star explodes as a SN leaving a compact object remnant that is later engulfed by the giant secondary. During this CEE event the NS/BH might spiral-in all the way into the core of the giant and merge with it. While compact object-core mergers progenitor systems where the stars are initially in relatively close orbits  experience a stable mass transfer event between the post-MS primary and the MS secondary (left panel; channel I), wider binary progenitors go through a CEE at this stage (right panel; channel II). 
}
\label{fig:EvolutionRoutes}
\end{center}
\end{figure*}

\begin{figure*}
\begin{center}
\includegraphics[width=0.95\textwidth]{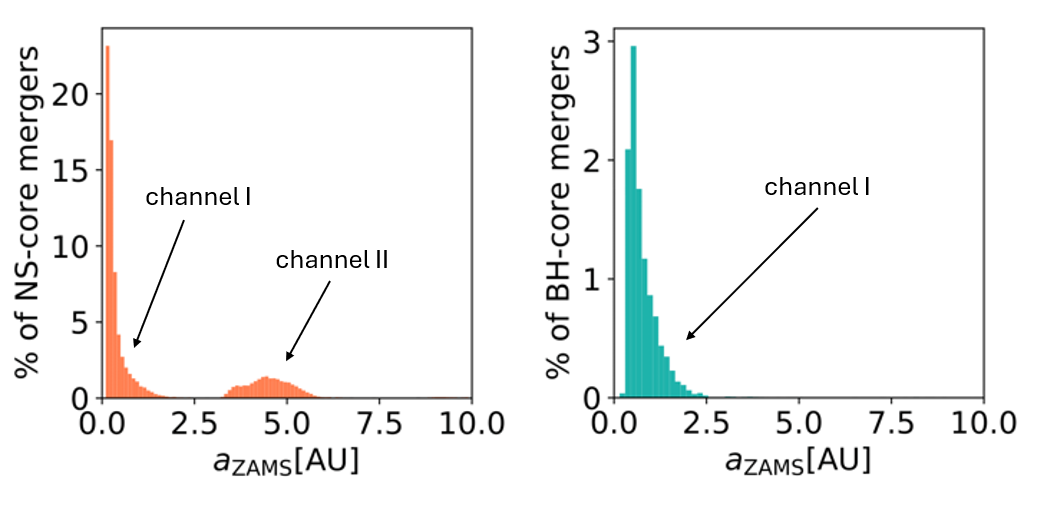}
\caption{Orbital separation distribution at the zero age MS of binary systems that will results in NS-core mergers (left panel; orange bins) and BH-core mergers (right panel; turquoise bins), expressed as percentages relative to the total number of systems where the compact object merges with the giant's core. Adapted from \cite{Grichener2023popSynth}. 
}
\label{fig:bimodality}
\end{center}
\end{figure*}

The mergers of NSs and BHs with cores of giant stars are believed to play a significant role in explaining luminous transients (section \ref{subsec:EMsignals}). Understanding the formation channels of binary systems that result in these mergers is important for determining where they are most likely to occur and identify precursor signatures that might assist in their detection. \cite{Grichener2023popSynth} performs binary population synthesis to study the possible evolutionary routes of binary systems towards NS/BH-core mergers and predict their rates. They identify two distinct evolution channels leading to NS-core mergers (Fig. \ref{fig:EvolutionRoutes}), which correspond to different regimes in the initial orbital separation distribution of the progenitor binary system (left panel of Fig. \ref{fig:bimodality}; orange bins). Only the main evolution channel (left panel of Fig. \ref{fig:EvolutionRoutes}) can lead to the merger of a BH with the core of a giant star, which is associated with the single mode in the right panel of Fig. \ref{fig:bimodality} (turquoise bins). Fig. \ref{fig:bimodality} shows the initial orbital separation distribution of system that results in NS/BH core mergers for a population synthesis model with a common envelope efficiency parameter \citep{LivioSoker1988CEE} of $\alpha_{\rm CE}=0.5$ and solar metallicity, but the trends shown in this plot are robust for all the studied  parameters (see \citealt{Grichener2023popSynth} for more details). 

The most dominant channel (left panel of Fig. \ref{fig:EvolutionRoutes}), which was previously discussed by \cite{Chevalier2012IIn} and \cite{Schroderetal2020mergers}, begins with a binary system that is composed of two massive stars relatively close to one another (initial orbital separation of $a_{\rm ZAMS} \lesssim 1.3 \; \rm au$). The initially more massive star (hereafter referred to as the \textit{primary star}) evolves past the main sequence (MS), fills its Roche lobe and begins transferring mass to its lighter MS companion (hereafter referred to as the \textit{secondary star}). This dynamically stable mass transfer episode might lead to the rejuvenation of the secondary star (e.g.,  \citealt{Hellings1984Rejuvenation}; \citealt{Renzoetal2023Rejuvenation}; \citealt{Waggetal2024}), modifying the core-envelope boundary region due to the accreted mass and decreasing significantly the envelope binding energy for the rest of the evolution.  The mass transfer remains stable and the primary eventually gets stripped off its envelope due to the mass lost to the secondary accretor (e.g., \citealt{KippenhahnWeigert1967stellarEvolution}; \citealt{Gotbergetal2017binaries}; \citealt{Gotbergetal2018binaries}) or through stellar winds (e.g., \citealt{Maederetal1981winds}. See \citealt{Beasoretal2022winds} for a different view), leaving a naked core that explodes as a stripped-envelope supernova (SN; e.g., \citealt{Podsiadlowskietal1992}; \citealt{Laplaceetal2021SN}; \citealt{Vartanyanetal2021SN}), observationally defined as type IIb or Ib/Ic, and produces an NS or a BH remnant. 

In cases where the binary system remains bound after the SN explosion, the expansion of the secondary star during its subsequent evolution can lead to a CEE phase that drastically reduces the orbital separation of the system due to dynamical friction (e.g., \citealt{Ostriker1999DynamicalFriction}; \citealt{Desjacquesetal2022DynamicalFriction}). The gravitational interaction between the moving compact object and the gaseous envelope that surrounds it creates density wakes that further interact with the compact object and bring it closer to the core, and its location is determined by drag forces applied by the surrounding envelope gas.  To date, there is no good analytic prescription for a motion of a star inside the shared envelope during CEE due to dynamical friction, and most population synthesis codes use the alpha-energy formalism (e.g., \citealt{vandenHeuvel1976CEE}; \citealt{Webbink1984CEE}; \citealt{LivioSoker1988CEE}) to find orbital separation at the end of CEE.

If the compact object manages to get close enough to the giant's core before the entire envelope is ejected then it will merge with it, resulting in a violent transient event. In cases where the tidal disruption radius of the core $r_{\rm t}$ exceeds the core’s actual radius $R_{\rm core}$, the core is tidally disrupted by the compact object’s immense gravity, preventing it from ever reaching the core's surface. The material of the core is dynamically stripped by the tidal torques, transporting angular momentum and forming an accretion disk around the compact object (e.g., \citealt{GrichenerSoker2023W49B}; \citealt{Eversonetal2024TZOs}). The accretion disk launches jets that interact with their environment by depositing part of the kinetic energy in the surrounding gas, which is then converted to radiation, powering a bright transient with observational signatures that could resemble rare SN explosions or similar transients (See section \ref{subsec:EMsignals}). The matter of the accretion disk could also experience thermonuclear outbursts (\citealt{GrichenerSoker2023W49B}; see section \ref{subsubsec:Thermonuclear}). 

Alternatively, if the tidal disruption radius lies within the core of the secondary star and in-spiral is fast enough, the compact object would enter while the core still possesses its original structure. In this case the mass accretion could occur either through an accretion disk formed by vast accretion onto the compact object as it spirals through the core, leading to a bright transient (e.g., \citealt{SokerGilkisiPTF14hls}; \citealt{Sokeretal2019FBOTs}) or be quasi-spherical powering a Thorne-Zytkow object (TZO) in case of an NS primary (e.g.,  \citealt{ThorneZytkow1975TZOs}; \citealt{ThorneZytkow1977TZOs}. See \citealt{OGradyetal2024TZOs} for a recent review) or a quasi-star in case of a BH primary (e.g., \citealt{Begelmanetal2006quasiStars}; \citealt{Begelmanetal2008quasiStars}; \citealt{Balletal2011quasiStars}), depending on the degree to which the core was spun-up by the compact object during CEE. If the specific angular momentum of the core material, $j_{\rm core}$, is larger than the angular momentum needed to maintain an orbit at the innermost stable circular orbit around a compact object, $j_{\rm isco}$, then an accretion disk will immediately form and a stable TZO configuration will be prevented, as found in the stellar evolution and hydrodynamical simulations analyzed by \cite{Eversonetal2024TZOs}. \cite{Nathanieletal2024TZOs} implement the condition of the tidal disruption radius in population synthesis data of massive binaries to explore a wider parameter space and determine which systems will avoid disruption and form classical TZOs. 

If the entire envelope is ejected before the compact object reaches the proximity of the core then they will avoid merging, and the stripped core might explode in a second stripped-envelope SN, leaving a second compact object behind. In cases where the compact objects in the double compact object (DCO) binary system remain close enough after the natal-kick induced by the SN explosion, they might merge within Hubble time producing GWs (e.g., \citealt{TutukovYungelSon1973binaries}; \citealt{Chevalier2012IIn}; \citealt{VignaGomezetal2018DNSs}; \citealt{Ablimitetal2018WDCO}; \citealt{Grichener2023popSynth}) and electromagnetic transients (e.g., \citealt{Janiuketal2013GRBs}; \citealt{Abbottetal2017bKilonova}) that are detectable with current facilities. Table \ref{table:DifferentOutcomes} summarizes the main evolutionary outcomes of a giant star - compact object CEE system.  

Binary systems with stars that begin relatively close to one another tend to produce NS-core mergers through a stable mass transfer episode between the post-main-sequence (post-MS) primary and the main-sequence (MS) secondary (stage 2 in channel I; left panel of Fig. \ref{fig:EvolutionRoutes}). Systems with larger initial separations ($3 \rm \; au  \lesssim a_{\rm ZAMS} \lesssim 7 \; \rm au$), however, might be too far apart for the initiation of a CEE evolution phase between the NS and the post-MS secondary after the first SN explosion. In this case, an earlier CEE that occurs while the secondary star is still on the MS and the primary star is more evolved could bring the core of the post-MS primary and the MS secondary closer together before the post-MS primary is entirely stripped of its envelope, as shown in the right panel of Fig. \ref{fig:EvolutionRoutes} (stage 3). This evolutionary route is less common, but still leads to a non-negligible number of NS-core merger events (up to $35 \%$ for commonly used model parameters, see \citealt{Grichener2023popSynth} for more details).  

\begin{table*}
\begin{center}
\hspace{-1.5cm}
\setlength\tabcolsep{3 pt}
\begin{tabular}{||l l ||}
 \hline
 \makecell{Orbital separation regime} & \makecell{Outcome}  \\ 
 \hline
  \makecell{$a_{\rm f}>R_{\rm core}$ and $a_{\rm f}>r_{\rm t}$  } & \makecell{Binary compact object system}  \\ 
 \hline
 \makecell{$R_{\rm core}<a_{\rm f} \lesssim r_{\rm t}$} & \makecell{Tidal-disruption induced merger}   \\ 
 \hline
 \makecell{$a_{\rm f}<R_{\rm core}$ and $r_{\rm t}<R_{\rm core}$ and $j_{\rm core}>j_{\rm isco}$ } & \makecell{Accretion-induced merger}     \\ 
 \hline
 \makecell{$a_{\rm f}<R_{\rm core}$ and $r_{\rm t}<R_{\rm core}$ and $j_{\rm core}<j_{\rm isco}$ } & \makecell{ TZO formation }  \\ 
 \hline
 \end{tabular} 
\end{center}
\caption{Possible outcomes of a giant star - compact object binary systems.}
{ \textbf{Definitions:} $a_{\rm f}$: orbital separation of the binary system at the end of CEE; $R_{\rm core}$: the radius of the core; $r_{\rm r}$: tidal disruption radius of the core; $j_{\rm core}$: specific angular momentum of the core material; $j_{\rm isco}$: angular momentum required to maintain an orbit at the innermost stable circular orbit around the compact object.
}
\centering
\label{table:DifferentOutcomes}
\end{table*}

\newpage

\section{Accretion and feedback physics during the merger}
\label{sec:physics}
 
One of the key processes that will determine the outcome of an NS/BH-core merger is the accretion rate by the compact object during CEE and the merger with the core. Therefore, properly modelling this accretion is crucial for understanding these mergers and the role they play in high energy astrophysics. 

Studies often use a fraction the Bondi-Hoyle-Lyttleton mass accretion rate (\citealt{HoyleLyttleton1939BHL}; \citealt{BondiHoyle1944BHL}) as an approximation for the rate through which the NS/BH will accrete mass from the envelope and core of the giant (e.g., \citealt{Papishetal2015Rprocess}; \citealt{GrichenerSoker2019rprocess}; \citealt{GrichenerSoker2021neutrinos}; \citealt{Hilleletal2022NJF3D}). This simplistic scheme assumes an accretor embedded in an infinite cloud of gas, where the trajectories of gas outflows are determined by the gravitational field produced by the accretor. \cite{HoyleLyttleton1939BHL} compare the gravitational energy of these outflows in a specific radius to their kinetic energy to understand whether they will be accreted.  \cite{BondiHoyle1944BHL} generalize this result by taking into account the pressure of the flowing matter. For a compact object that accretes mass from the core of a giant star this gives an accretion radius of 
\begin{equation}
R_{\rm BHL} =\frac{2GM_{\rm NS/BH}}{v_{\rm rel}^{2}+c_{\rm s}^{2}}, 
\label{eq:R_BHL}
\end{equation}
where $G$ is the gravitational constant, $M_{\rm NS/BH}$ is the mass of the compact object, $v_{\rm rel}$ is the velocity of the compact object relative to the core and $c_{\rm s}$ is the sound speed inside the core at the location of the compact object. The mass accretion rate is then  
\begin{equation}
\begin{split}
\dot{M}_{\rm BHL}\simeq \pi R_{\rm BHL}^{2} \rho_{\rm c} v_{\rm rel},
\end{split}
\label{eq:Mdot_BHL}
\end{equation}
where $\rho_{\rm c}$ is the density near the compact object. For a NS that merges with an evolved core of a massive star, this accretion rate can reach a value of several hundredths solar masses per second (e.g., \citealt{GrichenerSoker2019rprocess}).  
 
This analytical prescription does not take into account the possible effects of density gradients present in the envelope and core of the giant, tidal disruption of the core by the compact object, and ejection of gas by jet launching on the accretion rate. Therefore, conducting three dimensional hydrodynamical simulations to model this accretion is necessary for obtaining more accurate results. Hydrodynamical simulations that compute the accretion rates by a companion during different stages of CEE disagree on whether it constitutes a considerable fraction of the Bondi-Hoyle-Lyttleton mass accretion (e.g., \citealt{LivioSoker1986accretion}; \citealt{Chamandyetal2018accretion};  \citealt{LopezCamaraetal2020accretion}; \citealt{KashiMichaelis2022accretion}; \citealt{Prustetlal2024accretion}) or if it is much lower (e.g., \citealt{RickerTaamaccretion2012}; \citealt{MacLeodRamirezRuiz2015accreationA}; \citealt{MacLeodRamirezRuiz2015accreationB}). The different results are due to the high complexity in modelling CEE involving compact objects, that does not allow to include all relevant physical processes (such as magnetic fields, cooling mechanisms, gravitation and relativistic effects) and forces different studies to make different assumptions. 

One effect that would be crucial to explore in simulations in order to get a more realistic picture of the rate in which compact objects accrete mass during CEE is the role of radiation pressure on the accreted matter. In optically thick environments, as the accretion rate grows the in-flowing material heats-up and radiation pressure becomes so dominant that it can start counterbalancing gravity, inhibiting further accretion of material onto the accretor as the accretion rate reaches the Eddington limit, given by 
\begin{equation}
\begin{split}
\dot{M}_{\rm Edd,NS} = \frac{4\pi c R_{\rm NS}}{\kappa \eta} \quad \rm and \quad \dot{M}_{\rm Edd,BH} = \frac{4\pi G M_{\rm BH}}{\kappa \eta c}
\end{split}
\label{eq:Ledd}
\end{equation}
for a NS and BH accretors, respectively, where $c$ is the speed of light, $R_{\rm NS}$ is the radius of the NS, $\kappa$ is the typical opacity, $\eta$ is an efficiency factor and $M_{\rm BH}$ is the mass of the BH. For standard parameters, in both cases this accretion rate is on the order of $10^{-7}-10^{-6} M_{\rm \odot} \yr^{-1}$, much lower than the typical Bondi-Hoyle-Lyttleton accretion rate. While radiative cooling alone might not be able to increase the accretion rate by compact objects due to the high opacitties of stellar envelopes, convection could assist in carrying energy outwards to regions where radiative transfer becomes more efficient (e.g., \citealt{Gricheneretal2018Convection}; \citealt{WilsonNordhaus2019Convection}; \citealt{WilsonNordhaus2022Convection}.) To understand the effect of radiation on the accretion process, it is important to account for radiative transport in the simulations, as adiabatic hydrodynamic simulations of Bondi-Hoyle-Lyttleton mass accretion find that the infalling material creates a hydrostatic halo surrounding the compact object (e.g., \citealt{Ruffertetal1994accretion}; \citealt{Prustetlal2024accretion}). 

In the case of a compact object-core merger, the optically thick in-flowing gas is likely to trap the photons and carry them inwards, leading to adiabatic compression that increases the temperature of the flow. If it reaches the threshold for pair production, $T_{\rm pp}\simeq 1.2 \times 10^{10}\;\rm  K$ that corresponds to the total rest mass energy of the pair $E_{pp}=1.022 \MeV$, pairs of electrons $e^{-}$ and positrons $e^{+}$ can form and annihilate trough the production of photons that get trapped in the flow 
\begin{equation}
e^{-}+e^{+} \rightarrow \gamma + \gamma.
\label{eq:PP_gammas}
\end{equation}
In cases where the flow's temperature reaches values of $\approx 10^{11} \; K$, it becomes sufficiently high to provide the energy required for the production of (virtual) W and Z bosons that mediate weak interactions, and the newly formed $e^{-}e^{+}$ pairs could also annihilate by emitting neutrinos and anti-neutrinos
\begin{equation}
e^{-}+e^{+} \rightarrow \nu_{\rm e} + \bar{\nu}_{\rm e},
\label{eq:PP_nus}
\end{equation}
which stream out freely due to their low cross section of interaction with matter. The neutrinos escape with a fraction of the accretion energy on a timescale of mili-seconds \citep{GrichenerSoker2019rprocess}, drastically reducing the pressure near the compact object, hence allowing it to accrete at super-Eddington rates (e.g., \citealt{HouckChevalier1991neutrinoCooling}; \citealt{Chevalier1993neutrinoCooling}; \citealt{Fryeretal1996MergerDrivenExplosions}). \footnote{For BH-core mergers, it is more likely that the gas in-flowing into the BH will carry a larger part of the accretion energy beyond the horizon than the energy lost by neutrinos (e.g., \citealt{Pophametal1999BHs}).} \cite{Estebanetal2023neutrinos} find that detectable MeV neutrino signals are produced during the CEE of an NS with a star due to neutrino cooling that occurs when the NS accretes mass at these high rates. To fully explore the role of neutrino cooling in compact object-core mergers, it is important to note that it can be efficient at increasing the accretion rate by a compact object only in inner regions of the accretion flow where it overpowers viscous heating (e.g., \citealt{Metzgeretal2008neutrinoCooling}). In the outer regions viscous heating will drive wind mass loss, reducing the mass accretion rate reaching smaller radii (e.g., \citealt{StonePringle2001viscousHeating}; \citealt{Hawleyetal2002viscousHeating}; \citealt{YuanNarayan2014viscousHeating}).

If the accretion rate by the compact object is high enough, the specific angular momentum of the accreted gas could be larger than the specific angular momentum of a Keplerian orbit on the surface of the compact object, preventing its free fall onto the compact object and forming a centrifugally-supported accretion disk from the destroyed core material (e.g., \citealt{Soker2004accreationDisk}). Since the radius of the compact object ($\simeq 10^{6} \cm$) is much smaller than the typical orbital separation of the binary system at the onset of the merger ($\simeq 10^{8}-10^{10} \cm$), an accretion disk is expected to be formed. The friction between different segments of the disk cause inner segments to lose angular momentum and fall into the accretor releasing gravitational energy. The energy that is not radiated away could be re-processes to kinetic energy that would speed the rotation of the disk. Energy and angular momentum transport from the disk outwards could be performed through viscosity, magnetic torques and tidal interactions with the compact object (e.g., \citealt{PringleRees1972accreionDisk}; \citealt{ShakuraSunyaev1973accreionDisk}; \citealt{Dganietal1994accreionDisk}), and lead to the launching of disk winds or narrow jets. Even though jet physics is not fully understood, it is generally accepted that magnetic fields play a crucial role in their launching mechanism. \cite{BlandfordPayne1982jets} propose, for instance, that magnetic fields extract matter from the accretion disk to large distances, where they collimate that gas outflows into opposite narrow jets. The jets can then collide with the surrounding core matter, depositing their kinetic energy into the gas and remain choked (for more information on the propagation of jets in an external media see section \ref{subsec:neutrinos}). This kinetic energy is then converted to thermal energy and radiation, powering a bright transient. 

Jet launching could influence the accretion rate by the compact object as it spirals through the envelope and core of the giant via the negative jet feedback mechanism (see \citealt{Soker2016NJF} for a review). In this feedback cycle, jets that the accretion disk launches collide with and deposit their kinetic energy in their surroundings gas, expelling some of it away. Since the jets are powered by accretion of matter from the compact object's surroundings, by expelling the matter around them and leaving less gas to accrete, their power drops. The less powerful jets will then be able to expel less gas around them, until they encounter a fresh supply of new gas as the compact object keeps spiraling-in. \cite{Gricheneretal2021NJF} and \cite{Hilleletal2022CEJSN} find that this negative feedback cycle can lower the accretion rate by a factor of about 10 compared to the accretion rate originally assumed.

\section{R-process nucleosynthesis and enrichment}
\label{sec:rprocess}

The possible formation site(s) of r-process elements remain a highly debated subject (e.g., \citealt{Hotokezakaetal2018rprocess}; \citealt{Siegeletal2019collapsar}; \citealt{Jietal2019testForNSmergersRprocess}; \citealt{Banerjeeetal2020rprocess}; \citealt{Chenetal2024rprocess}; \citealt{Chenetal2024rprocessOther}; \citealt{Tsujimoto2024rprocess}). In 2017, LIGO detected the first GW signature from binary NS (BNS) mergers, commonly known as GW170817 \citep{Abbottetal2017aGW}. The electromagnetic counterpart that was associated with this GW event (kilonova AT2017gfo; \citealt{Abbottetal2017bKilonova}) showed traces of lanthanides in their equatorial outflow \citep{Kasenetal2017TheRprocessPaper}, hence confirming r-process nucleosynthesis in BNS mergers. Despite this groundbreaking progress, which was further strengthened by an additional discovery of a kilonova that might be associated with another BNS merger with \textit{JWST} \citep{Levanetal2024kilonova}, it is not clear whether BNS mergers are the only site of heavy element formation in the Universe. The long delay times until the merger and subsequent r-process nucleosynthesis, $\propto t^{-1}$ reaching tens of Gyrs (e.g., \citealt{Belczynskietal2018BNS}), might imply there are not enough mergers to account for heavy element formation in the early Universe (e.g., \citealt{Coteetla2019rprocess}; \citealt{Kobayashietal2023rprocess}. See \citealt{MaozNakar2024rprocess} and \citealt{Beniaminietal2024rprocess} for a different view). Moreover, these long delay times combined with the two natal-kicks that the binary system receives following each SN explosion (see left panel of Fig. \ref{fig:EvolutionRoutes}) might drive the BNS outside of ultra-faint dwarf (UFD) galaxies before they manage to merge, implying that they are not able to enrich them with heavy elements (e.g., \citealt{Bonettietal2019rprocess}). However, recent observations of the UFD galaxy Reticulum II show r-process enhanced gas (e.g., \citealt{Jietal2016RetII}; \citealt{Roedereretal2016RetII}; \citealt{Jietal2023RetII}; \citealt{Simonetal2023RetII}), highlighting the need for additional r-process nucleosynthesis sites (see \citealt{Beniaminietal2016rprocess} for a another view.)

\cite{Papishetal2015Rprocess} suggest that r-process nucleosynthesis can occur during the merger of an NS with the core of a giant star. As the NS spirals through the core of the giant, it might accrete mass at very high rates (see section \ref{sec:physics} for a discussion about the accretion rate). The vast accretion destroys the core forming a thick accretion disk around the NS that can self neutronize through electron captures due to the extremely high temperatures and densities in the disk, providing a neutron-rich environment where heavy elements can be formed via rapid neutron captures by isotopes that come from outer cooler layers of the disk. \cite{GrichenerSoker2019rprocess} showed that under their model assumptions the matter of the accretion disk could be neutron-rich enough for r-process nucleosynthesis if the merger occurs after helium core depletion, when the core is CO-rich. They perform one-dimensional stellar evolution simulations of several massive stars that engulf an NS in different stages of their evolution and mimic the effect of the NS on the giant by removing envelope mass from the location of the NS outwards. Using the Bondi-Hoyle-Lyttleton approximation (see equation \ref{eq:Mdot_BHL}), they compute the accretion rate of the NS as it merges with the core of the giant, and then apply it to estimate the typical temperatures and densities of the material in the disk within the analytical framework of a neutrino cooled accretion disk from \cite{Chevalier1996neutrinoCooling}
\begin{equation}
T \approx  5.5\times 10^{10}  
\left( \frac{\dot M}{0.05 M_\odot \s^{-1}} \right)^{0.11}  
\left(\frac{M_{\rm NS}}{1.4 M_\odot} \right)^{0.16} 
\left(\frac{r}{50 \km} \right)^{-0.47} \K .
\label{eq:temperature}
\end{equation}
and
 \begin{equation}
\rho \approx 6\times 10^{10}   
\left( \frac{\dot M}{0.05 M_\odot \s^{-1}} \right)^{0.84}  
\left(\frac{M_{\rm NS}}{1.4 M_\odot} \right)^{0.76} 
\left(\frac{r}{50 \km} \right)^{-2.29}
\left(\frac{\alpha}{0.01} \right)^{-1} \g \cm^{-3} , 
\label{eq:density}
\end{equation}
where $\dot M$ is the accretion rate by the NS during the merger, $M_{\rm NS}$ is the mass of the NS, $r$ is the radius of the accretion disk and $\alpha= \nu /c_s H$, where $\nu$ is the coefficient of the kinematic viscosity, $c_s$ is the sound speed and $H$ is the vertical scale height of the disk. These typical parameters lead to an electron fraction $Y_{\rm e} \lesssim 0.2$ (e.g., \citealt{Beloborodov2003rprocess}), implying the matter in the accretion disk is neutron-rich enough to produce Lanthanides \citep{JinSoker2024Rprocess}. Since the temperature across the accretion disk is not homogeneous, the inner hotter layers would reach the temperatures required for neutrino cooling (see section \ref{sec:physics}). \cite{GrichenerSoker2019rprocess} find that in cases where the giant star engulfs the NS earlier in the evolution (when the core of the star is less dense and composed mainly out of helium) the relatively low accretion rates lead to cooler disks where r-process nucleosynthesis cannot occur. In these cases we would also expect the synthesis of low amounts of $^{56}\text{Ni}$, in coincidence with lightcurves of LFBOTs (see section \ref{subsubsec:FBOTs}). 

\begin{figure}
\begin{center}
\vspace*{-7.10cm}
\hspace*{-0.95cm}
\includegraphics[width=1.0\textwidth]{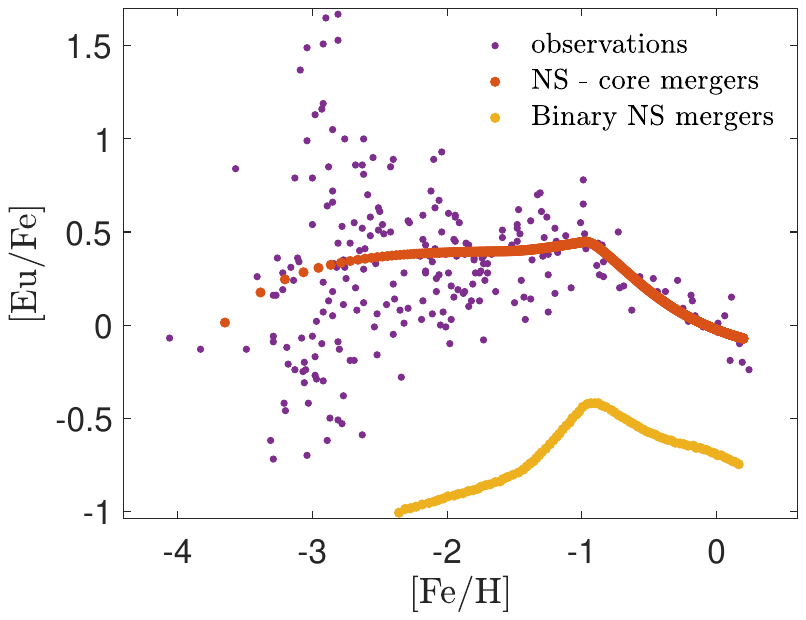}
\vspace*{-7.2cm}
\caption{Time Evolution of europium in the Milky-Way Galaxy due to the contribution of NS-core mergers (orange dots) and BNS mergers (yellow dots). The purple dots represent observations of individual stars. Adapted from \cite{Kobayashietal2020GCE} and \cite{Gricheneretal2022GCE}. 
}
\label{fig:EuOverFeScatter}
\end{center}
\end{figure}

Since the orbital shrinkage leading to mergers of NSs with giants' cores is driven by dynamical friction during CEE rather than GW irradiation, they occur on a dynamical timescale and have much shorter delay time from the formation of the binary compared to BNS mergers ($\simeq$ tens of Myrs; from the population synthesis data of \citealt{Grichener2023popSynth}; comparable to the lifetime of the NS progenitor). These shorter delay times offer an advantage in explaining the presence of heavy element in the early Universe. Moreover, NS-core mergers suffer from a single natal-kick induced by the SN that formed the NS, increasing the possibility that they will be retained and produce heavy elements in UFD galaxies \citep{GrichenerSoker2022RetII}. \cite{Grichener2023popSynth} computes the event rates of NS-core mergers with CO-rich cores and find it to be around 1 event per 1000 CCSNe for fiducial model parameters, which is comparable to the rate of BNS mergers inferred by LIGO \citep{Abbottetal2013LIGO}, and sufficient to account for a substantial fraction of observed r-process abundances. \cite{Gricheneretal2022GCE} use a Galactic Chemical Evolution model of the Milky-Way \citep{Kobayashietal2020GCE} and include the contribution of NS-core mergers to r-process nucleosynthesis. They study the time evolution of europium in the Galaxy and find that europium nucleosynthesis yields in NS-core mergers (orange dots in Fig. \ref{fig:EuOverFeScatter}) are within the confines of the observed values at earlier times and they can qualitatively reproduces the 'evolution knee' that follows, while BNS mergers (yellow dots in Fig. \ref{fig:EuOverFeScatter}) struggle to do so.

The previously discussed studies show NS-core mergers have the potential to explain r-process formation and enrichment in astrophysical domains where BNS mergers face difficulties \footnote{Other suggested astrophysical sites with potential to do so are outflows from MHD-driven supernovae; e.g. \citealt{Winteleretal2012MHDsn} and relativistic jets formed in collapsars; e.g., \citealt{Siegeletal2019collapsar}.}. However, three dimensional magneto-hydrodynamical simulations are required to fully explore this potential (see also section \ref{sec:future}). To properly obtain the neutron richness of that accretion disk matter, these simulations should also account for the possible re-absorption of neutrinos that are radiated as the disk cools down (see \citealt{Issaetal2024Collapsar} for recent Neutrino-Gravitational Magneto Hydrodynamical simulations of collapsars).     

\newpage

\section{Observational signatures}
\label{sec:observations}

\subsection{Electromagnetic signatures}
\label{subsec:EMsignals}

\subsubsection{Merger-driven lightcurves properties}
\label{subsubsec:GeneralObs}

As discussed in section \ref{sec:physics}, when an NS or a BH merges with the core of a giant  the angular momentum of the merger can lead to the formation of a neutrino cooled accretion disk (e.g., \citealt{Chevalier1996neutrinoCooling}; \citealt{GrichenerSoker2019rprocess}). A fraction of the accretion energy that does not escape the systems through neutrinos can be used to launch disk winds or narrow jets. The jets interact with their environment by depositing their kinetic energy in the surrounding envelope gas generating shocks that propagate through the unbound envelope that forms a dense circumstellar medium (CSM) around the merger. In general, as the shocked ejecta collides with the CSM, it transfers its kinetic energy to the CSM gas, converting it to thermal energy and radiation that might stream out freely from the merger zone or get trapped and re-processed until further expansion. Therefore, the distribution of the CSM has a key role in shaping the light curve that results from these merger driven explosions (e.g., \citealt{Chevalier2012IIn}; \citealt{Sokeretal2019FBOTs}; \citealt{Schroderetal2020mergers}). 

While three-dimensional hydrodynamical simulations that follow the the outflows and outgoing radiation during the merger of the compact object with the core are scarce, simulating the final CEE stages of a compact object and a giant star has value in building intuition. \cite{Schroderetal2020mergers} model the in-spiral of a BH inside a giant and calculate the lightcurves expected prior to the merger of the BH with the core accounting for the effects of CSM interaction with a spherically generated shock instead of jets. They continue the simulations until the orbital separation of the binary is about $a = 100 R_{\rm \odot}$, assuming the subsequent merger is expected to occur rapidly on a timescale of less then a year. They then analyze the ejection of matter during the final stage of CEE and find the resultant CSM distribution . We note that by not modelling the accretion onto the BH, their resultant CSM might be more massive than expected in such a merger. For computational simplicity, they derive a spherically symmetric density distribution of the ejecta and model the CSM interactions using a one-dimensional radiative hydrodynamics code. 

\begin{figure}
\begin{center}
\hspace*{-0.95cm}
\includegraphics[width=0.5\textwidth]{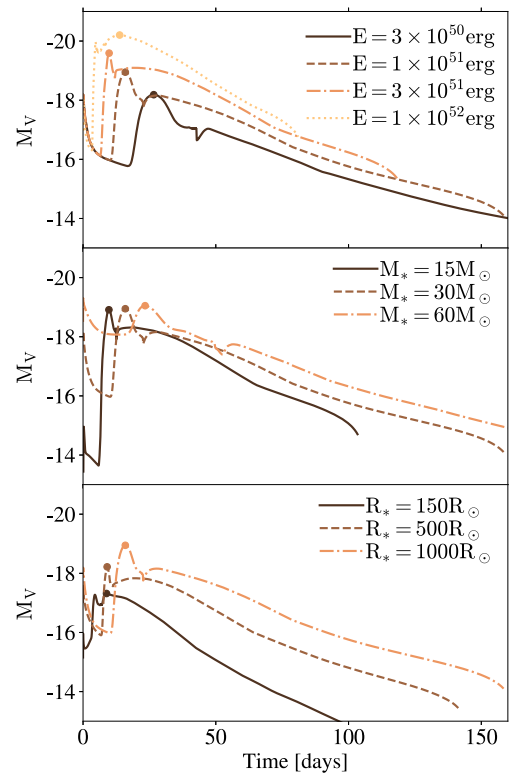}
\caption{Modeled lightcurves of merger driven explosions computed during the final CEE stages prior to the merger of a giant and a BH with mass ratio of $q \equiv M_{\rm *} \slash M_{\rm BH}=0.1$, taken from \cite{Schroderetal2020mergers} with permission from A. Vigna-Gomez. The top, middle and bottom panels show variation in explosion energy E, mass of the giant $M_{\rm *}$  and radius of the giant $R_{\rm *}$ at the onset of CEE, respectively, from the 'fiducial' model of $E=10^{51} \rm ergs$, $M_{\rm *} = 30M_{\rm \odot}$ and $R_{\rm *} = 1000R_{\rm \odot}$. 
}
\label{fig:MergersLightCurves}
\end{center}
\end{figure}
As can be seen in Fig \ref{fig:MergersLightCurves}, \cite{Schroderetal2020mergers} find that at early times the lightcurves are mostly dominated by CSM interactions. As in other models of dense CSM close to the giant star, they find fast rising high luminosities that reach plateau over a period of hundreds of days. \footnote{In the case of an extended CSM the decay of the lightcurve can happen over a period of thousands of days (e.g., \citealt{Smith2017}).} Typical models have higher luminosities than type IIp SNe, but still lower than in the case of super-luminous SNe. The top panel of Fig. \ref{fig:MergersLightCurves} shows that more energetic mergers result in shorter, more luminous transients with less prominent effect of CSM interactions on the lightcurve as the larger photospheres of the explosive transients are further out from the CSM edge. The lightcurve of the most energetic transient ($E=10^{52}$ ergs) resembles the shape of a type IIp SN \citep{Schroderetal2020mergers}. Higher giant masses (middle panel of Fig. \ref{fig:MergersLightCurves}) result in lower ejecta velocities that lead to longer transients. Different giant radii (bottom panel of Fig. \ref{fig:MergersLightCurves}) affect the transients mainly by forming CSMs of different extents. Smaller giants lead to more compact CSMs that can affect the lightcurve through interactions at earlier times. Such explosive transients could be detected by current facilities, such as \textit{ZTF} \citep{Bellmetal2019ZTF} and \textit{ASAS-SN} \citep{Kochaneketal2017ASASSN}, as well as with future programs like \textit{LSST}. The CSM interactions involved in shaping the lightcurves generate X-ray radiation that could be followed with \textit{XMM-Newton} \citep{Jansenetal2001XMMnweton} and \textit{SWIFT} \citealt{Gehrelsetal2004SWIFT}.

Including the effect of jets in simulating the in-spiral of the compact object inside the giant and modeling their interaction with the ejecta and CSM might yield different results for the lightcurves. Modeled jet-shaped lightcurves exhibit higher luminosity peaks compared to a spherical explosion as a result of the extra energy source and relativistic beaming, and more abrupt declines due to a faster polar ejecta  (e.g., \citealt{KaplanSoker2020lightcurve}). The aspherical nature of jet launching would lead to a highly asymmetrical CSM, variability in the optical and X-ray emissions and a bumpy lightcurve (see \citealt{Schreieretal2021CEJSN} for lightcurves powered by jets that are launched when an NS crosses the envelope of a giant star in an eccentric orbit). The negative jet feedback mechanism (e.g., \citealt{Gricheneretal2021NJF}; \citealt{Hilleletal2022NJF3D}; see section \ref{sec:physics}) would lead to changes in the accretion rate by the compact object and therefore in the jet luminosity, contributing to the variability in the emission. The large amount of mass expelled due to jet launching during the merger would expand faster than in a case with no jets, and could lead to enhanced dust formation with strong IR radiation that could be observed with \textit{JWST} \citep{Gardneretal2006JWST} and the upcoming \textit{Nancy Grace Roman Telescope}. In cases where the merger occurs after the ejection of most of the envelope, the jets could stream out freely and give rise to a LGRB in the wavelength range of \textit{SWIFT}(\citealt{Gehrelsetal2004SWIFT}; see section \ref{subsubsec:GRBs}). In this case the ejecta would exhibit low amounts of hydrogen, as in Type Ibn (e.g., \citealt{Pastorelloetal2007ibn}) and Icn SN (e.g., \citealt{GalYametal2023ICN}).Further study and modelling are needed to determine the role jets would have in shaping the lightcurves of NS/BH-core mergers, and give prediction of observational signatures to search for them in existent data and future surveys.

\subsubsection{Core collapse supernovae}
\label{subsubsec:SNe}

Several studies have been exploring the potential of NS/BH core mergers to explain different transients that are observationally classified as sub-classes of CCSN. \cite{Chevalier2012IIn} argues that the merger of an NS or a BH with the core of a giant star during CEE might account for type IIn SNe events \citep{Schlegel1990IIn}, which are believed to have CSM around them due to the narrow lines in their spectra. Their velocity, mass and timescale are in rough agreement with the mass lost during CEE. Their optical luminosities could arise from the interaction of the CSM with the shock waves generated during the merger. Most luminous SNe IIn tend to occur in environments with low metallicities (e.g., \citealt{Neilletal2011IIbn}) which is also the regime where more massive stars can become giants due to the lack in stellar winds and engulf a compact companion. \cite{Schroderetal2020mergers} compare their modeled lightcurves for merger driven explosions (section \ref{subsubsec:GeneralObs}) with the observed lightcurve of the prototypical type IIn SNe SN1998S (e.g., \citealt{Shivversetal2015IIn}), and find a good compatibility in terms of the overall lightcurve shape, luminosity and duration \footnote{\cite{Schroderetal2020mergers} also find that their lightcurves could explain the observations of SN1979c that is classified as Type IIL SN.}. The presence of a close and asymmetric CSM around this SN could be explained by the fact that mass loss occurs a short time (few years) prior to the explosion (e.g., \citealt{Ofeketal2013binaries}), and is expected to be denser near the equatorial plane of the binary \citep{Chevalier2012IIn}.

Based on \cite{Chevalier2012IIn}, \cite{Dongetal2021mergerTriggeredCCSN} attributes the luminous radio transient VT J121001+495647, associated with the soft X-ray burst LGRB 140814A, to a merger triggered CCSN where relativistic jets were launched during the explosion. The radio emission arises from synchrotron radiation produced by shocks that are formed as jets interact with their surroundings. An NS/BH-core merger progenitor is consistent with the dense extended asymmetric CSM formed around the transient (see section \ref{subsubsec:GeneralObs} and section \ref{subsubsec:FBOTs}) and the pre-explosion eruptive mass loss episodes could originate in earlier mass transfer phases within the binary system.  

\cite{BarkovKomissarov2011hypernova} ties NS-core mergers to hypernova explosions. They propose that the high mass accretion rates during the merger can lead to the generation of strong magnetic fields through differential rotation energizing a very luminous Type II SN with a high plateau followed by a steep decline. The hydrogen features come from the giant's envelope, and the sharp drop in the brightness is due to low amounts of  $^{56}\text{Ni}$ synthesized is such an explosion that occur in relatively low densities and the fast spin-down of the NS. We note that the spinning of the NS in this scenario, and hence its potential to produce a magnetically driven explosion during the merger, is very sensitive to the specific angular momentum of the matter accreted by the NS itself. If the vast accretion leads the NS to collapse into a BH before the magnetically driven explosion manages to occur, then the relativistic jets launched due to the accretion by the BH will most likely remain choked in the inflated envelope of the companion (e.g., \citealt{GrichenerSoker2021neutrinos}), leading to a superluminous Type II SN. \citealt{Ablimitetal2022WDs} suggest that a similar transient can arise from the collapse of a white-dwarf to a NS or a BH as it mergers with the core of a giant star during CEE.  

More recently, an NS-core merger origin (\citealt{SokerGilkisiPTF14hls}; \citealt{Gofmanetal2019iPTF14hls}) was suggested for the peculiar type IIp SN iPTF14hls \citep{Arcavietal2017iptf14hls} and iPTF14hls-like transients, such as SN2020faa \citep{Yangetal2021SN2020faa}. Their unusual properties include an order of magnitude slower time evolution compared to a regular type IIp SN, at least five peaks in the lightcurve and a rapidly expanding CSM. Mass accretion by an NS as it passes the periastron in its eccentric orbit around a giant prior to the onset of CEE could explain observed pre-explosion outbursts (e.g., \citealt{Schreieretal2021CEJSN}). The jets launched while the NS is spiraling-in inside the envelope of the giant star lead to the ejection of the CSM around the transient. When the NS mergers with the helium core it launches more luminous jets due to higher accretion rates that power the explosion itself. Interactions of the jets with the ejected SN matter and late episodes of jet-launching might explain the multiple peaks in the lightcurve (e.g., \citealt{KaplanSoker2020lateJets}). Further accretion of fallback gas can prolong the evolution of the transient.

\subsubsection{Thermonuclear explosion}
\label{subsubsec:Thermonuclear}

\begin{figure}
\begin{center}
\hspace*{1.3cm}
\includegraphics[width=0.7\textwidth]{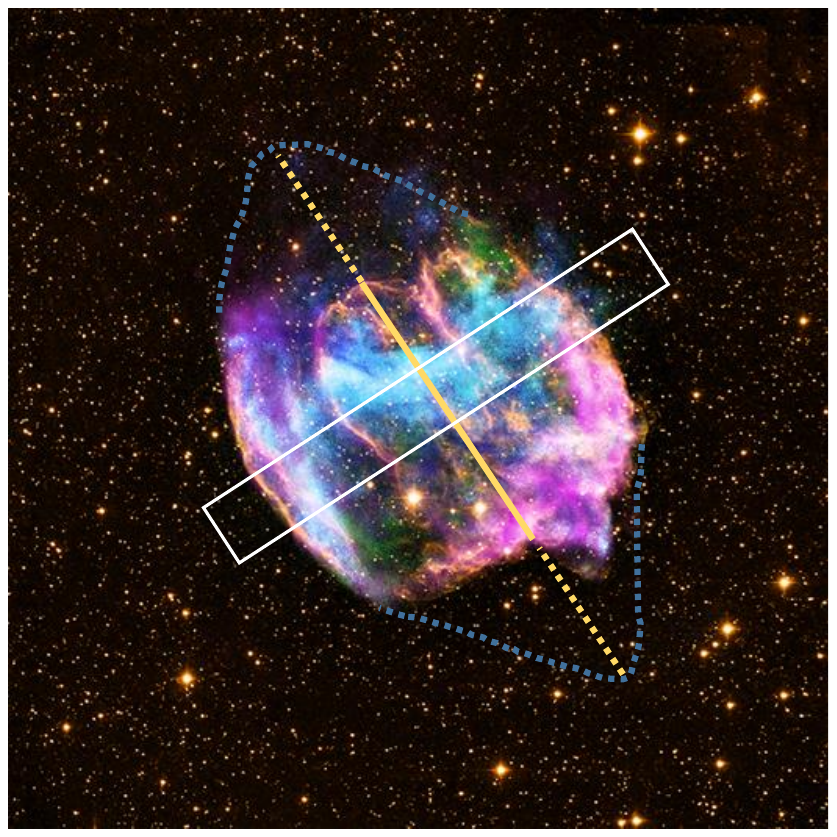}
\vspace*{-6.5cm}
\caption{Image of supernova remnant W49B based on \citealt{Lopezetal2013W49B}, with additional drawings  outlining the morphology discussed in \citealt{BearSoker2017W49B}. The yellow line marks the symmetry axis and the dotted cyan lines encompass the two potential past lobes. The white rectangle indicate the approximate equatorial plane of the NS-core binary system. Taken from \citealt{GrichenerSoker2023W49B}.}
\label{fig:W49B}
 \end{center}
 \end{figure}

In cases where the tidal disruption radius of the core is outside of the core (third row of table \ref{table:DifferentOutcomes}), the NS/BH will exert tidal forces that disrupt the core before it reaches the core's surface. In this case, the accretion disk formed around the compact object could experience thermonuclear outbursts, powering a bright transient (e.g., \citealt{Ginatetal2020GWs}; \citealt{GrichenerSoker2023W49B}), with some similarities to the disruption of a white dwarf by a compact object (e.g., \citealt{Fryeretal1999WD}; \citealt{Zenatietal2019WD}; \citealt{Zenatietal2020WD}). \citealt{GrichenerSoker2023W49B} propose that this kind of NS-core thermonuclear merger could be the origin of the enigmatic SN remnant W49B. The ignition of thermonuclear burning in the non-axisymmetric disk could account for the highly uneven distribution of metals, and its peculiar morphology (see Fig \ref{fig:W49B}) could be attributed to jet-ejecta interactions \citep{Akashietal2018W49B}. If the nuclear burning and jet-launching occurs relatively early in the tidal disruption driven merger, the disk will not be fully symmetric and the disrupted material will be concentrated in two spiral tails, leading to its ejection in an elongated structure. 


\subsubsection{Luminous fast blue optical transients}
\label{subsubsec:FBOTs}

Fast blue optical transients ( e.g., \citealt{Droutetal2014FBOTs}; \citealt{Arcavietal2016FBOTs}) are rapidly evolving transients with luminosities that can reassemble SN explosions. The most luminous subclass of fast blue optical transients is commonly refered to as LFBOTs, or AT2018cow-like transients, named after the most studied object in this group (e.g., \citealt{Marguttietal2019LFBOTs}; \citealt{Coppejansetal2020LFBOTs}; \citealt{Hoetal2020LFBOTs}; \citealt{Hoetal2023LFBOTs}). AT2018cow has an optical rise time of a few days and a peak luminosity $L_{\rm opt} \simeq 4 \times 10^{44}$ erg $\rm s^{-1}$ that then declines as $L_{\rm opt} \propto t^{-2}$. The optical light curve does not display a second peak, implying a low amount of $^{56}\text{Ni}$ in the ejecta. 

The initial spectrum of AT2018cow indicates expansion of the ejecta in velocities close to $10 \%$ of the speed of light, while in later spectra we can see a decrease of about a factor of ten in ejecta velocities (e.g., \citealt{Marguttietal2019LFBOTs}). This decrease is accompanied by a steep decay in the X-ray emission at a rate faster than $L_{\rm X} \propto t^{-4}$, but with no change in the ejecta temperature. The variability in the optical and X-ray emissions can be explained if a compact source is powering the event, and the large range in the velocities points towards an highly aspherical ejecta that surrounds it. 

LFBOTs tend to be associated with low-mass, low metallicity star forming galaxies (e.g., \citealt{Hoetal2020LFBOTs}; \citealt{Lymanetal2020LFBOTs}; \citealt{Coppejansetal2020LFBOTs}), similar to host galaxies of other explosive transients such as LGRBs, favoring very massive progenitors that require low metallcities to form. The CSM around LFBOTs is found to be hydrogen depleted and extended compared to predictions from massive star winds  \citep{FoxSmith2019LFBOTs}. The interaction between the fast polar ejecta and the CSM gives rise to radio and synchrotron emission. 

Since their discovery, many models were proposed to explain the origin of LFBOTs (e.g., \citealt{Perleyetal2019LFBOTs}; \citealt{FoxSmith2019LFBOTs}; \citealt{Leungetal2020LFBOTs}; \citealt{Kremeretal2021LFBOTs};  \citealt{Xiangetal2021FBOTs}; \citealt{Gottliebetal2022FBOTs}; see \citealt{Metzger2022FBOTs} for a list of scenarios and a comparison between them). \cite{Metzger2022FBOTs} concludes that the biggest challenge to any model that tries to explain LFBOTs is to account for the simultaneous presence of an energetic compact object and a massive and radially extended CSM surrounding the explosion. Models that are based on CSM interactions alone (e.g., \citealt{FoxSmith2019LFBOTs}) and do not involve a compact object as engine have difficulty in explaining the time-variable non-thermal emission that likely comes from a compact object ejecta. CCSN models (e.g., \citealt{Perleyetal2019LFBOTs}), on the other hand, have trouble reproducing the asymmetric ejecta that is shaped by the wide range of outflow velocities. An initially failed SN event that forms a BH and ejects mass through disk winds/jets might be able to explain the asymmetry of the ejecta, but will have challenges accounting for the presence of massive CSM around the LFBOTs. 

Several studies suggested that LFBOTs could originate in the mergers of NSs or BHs with helium cores of giant stars (\citealt{Sokeretal2019FBOTs};  \citealt{Soker2022FBOTs}; \citealt{Metzger2022FBOTs}; \citealt{CohenSoker2023FBOTs}; \citealt{TunaMetzger2023CBDs}). While the classical prompt NS/BH-core merger scenario during \citep{Sokeretal2019FBOTs} or shortly after \citep{Soker2022FBOTs} CEE can explain both the presence of a powering compact object and a massive CSM around the LFBOT, it has a problem accounting for the lack of hydrogen in the CSM, as hydrogen-rich debris of the common envelope would probably give rise to hydrogen features in the transient spectra. \cite{Metzger2022FBOTs} suggest that LFBOTs originate in the merger of an NS or a BH with the naked core of the giant. \footnote{If sufficiently massive and luminous, the naked core might appear as an hydrogen-poor classical Wolf-Rayet star. See \citealt{Shenar2024WRs} for a recent review.} The NS/BH-core binary system could be formed following the ejection of the common envelope at the end of the CEE phase (stage 7 in channel I, left panel of Fig. \ref{fig:EvolutionRoutes}) or following a stable mass transfer episode that leads to Roche lobe Overflow (RLOF) and could also power fast radio bursts at an early stage (e.g., \citealt{Sridharetal2021FRBs}). The merger is delayed by about $\simeq 100$ yr compared to the mass transfer event and occurs due to the interaction of the surviving binary system with a circumbinary disk that is formed around the system (for a population synthesis study that explores the conditions for the formation of circumbinary disks in post common envelope binary systems with compact objects see \citealt{Ungeretal2024CBDs}). Loss of angular momentum to the disk through viscous torques causes the orbit of the binary to shrink until the compact object merges with the core when most of the envelope matter is far from the location of the merger, explaining the small amounts of hydrogen in the CSM. The dense CSM can come from mass that the core loses at the beginning of the merger and from winds due to the photo-evaporation of the circumbinary disk. The disk-wind outflow launched during the merger could explain both the asymmetry of the ejecta with the different velocity components that come from the non-spherical nature of these outflows, and the low amounts of $^{56}\text{Ni}$ synthesized in the inner regions of the disk where the accretion rates and temperatures are high enough. Since the compact object merges with the helium core of the giant, we do not expect self-neutronization of the disk and r-process nucleosynthesis in these events.

\begin{figure}
\begin{center}
\hspace*{-0.95cm}
\includegraphics[width=0.7\textwidth]{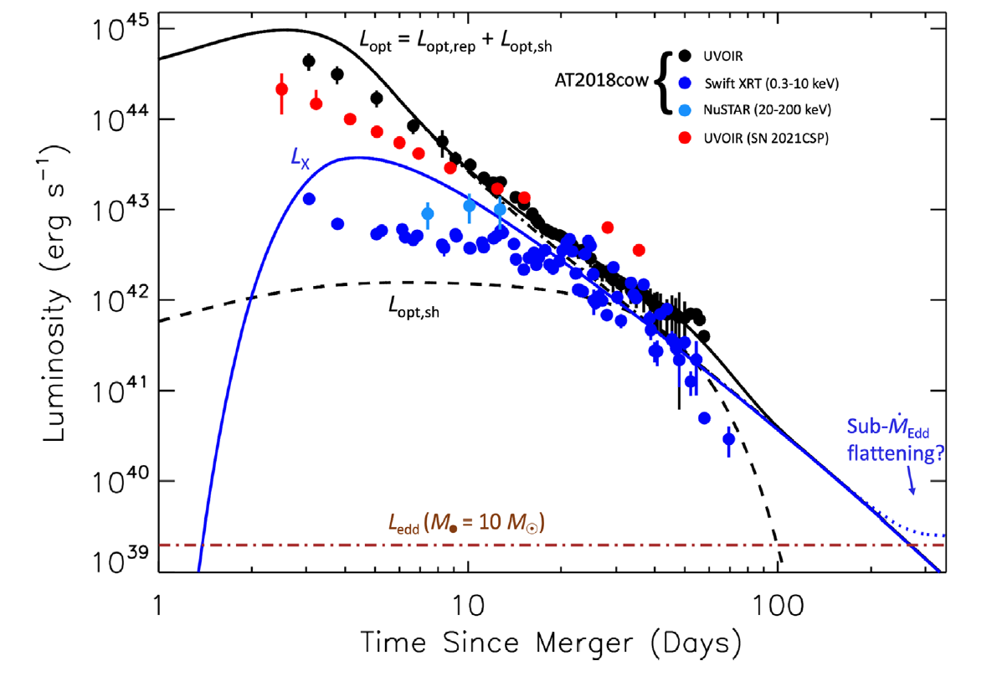}
\caption{Optical and X-ray light curve model of the transient that results from the merger of a $10M_{\rm \odot}$ an exposed core with a $10M_{\rm \odot}$ BH, taken from \cite{Metzger2022FBOTs} with permission from B. Metzger. The total optical luminosity $L_{\rm opt}$, represented by the solid black line, has a component from reprocessing the X-rays that are emitted in the inner disk outflow by the fast disk-wind ejecta ($L_{\rm opt,rep}$; black dotted-dashed line) and a component from the interaction of the CSM with the shocked disk-wind ejecta ($L_{\rm opt,sh}$; black dashed line). The solid dark blue line is the X-ray luminosity $L_{\rm X}$, and the dotted blue line represents a possible flattening of the X-ray lightcurve when the mass accretion onto the BH powers a transient that reaches the Eddington value  ($L_{\rm Edd}$; brown dotted-dashed line). The black, dark blue and light blue dots are observations of AT2018cow from \cite{Marguttietal2019LFBOTs} at optical, soft X-ray and Hard X-ray energies, respectively, while the red dots are optical observations of SN 2021csp classified as Type Icn SN. \citep{Perleyetal2022SN2021csp}.
}
\label{fig:LFBOTsLightCurves}
\end{center}
\end{figure}

Fig. \ref{fig:LFBOTsLightCurves} (taken from \citealt{Metzger2022FBOTs}) shows a good compatibility between the total optical (solid black line) and X-ray (solid dark blue line) lightcurves powered by a naked core-BH delayed merger model of an equal mass binary of $10M_{\rm \odot}$ and the observed AT2018cow energies from \cite{Marguttietal2019LFBOTs} in the respective bands (black, dark blue and light blue dots for optical, soft X-ray and hard X-ray, respectively). The optical luminosity is powered by both the shock interaction between the ejecta and CSM (black dashed line) and re-processing of the X-ray emission from the disk (black dotted-dashed line). The luminosity of the X-rays that manage to escape and avoid being re-processed $L_{\rm X}$ is presented by the solid dark blue line, that might flatten (dotted blue line) as the BH mass accretion rate reaches the Eddington value (the Eddington luminosity is presented by the brown dotted-dashed line) as seen in X-ray binaries. While the steepening of the X-ray light curve around a month is not obtained by this toy-model, it could be potentially explained by a reduced mass accretion rate onto the BH due to expansion of the disk. 

\subsubsection{Long gamma ray bursts}
\label{subsubsec:GRBs}
`
The mergers of BHs with cores of giant stars have been previously proposed as possible progenitors of  LGRBs (e.g., \citealt{FryerWoosley1998HeStarBHmerger}; \citealt{ZhangFryer2001HeStarBHmerger}). The rapid accretion of matter from the rotating core through an accretion disk formed around the BH leads to launching of relativistic jets that could energize the LGRB explosions. The gamma-rays are produced by synchrotron radiation that is emitted as the jets interact with their surroundings, as in the case of collapsars that come from single stars (e.g., \citealt{MacFadyenWoosley1999collapsars}). We note, however, that when the merger occurs inside the opaque envelope of the giant the jets are likely to get choked by the envelope gas (e.g., \citealt{Papishetal2015Rprocess}; \citealt{GrichenerSoker2021neutrinos}) and trap the gamma-rays. Therefore, it is more likely that such LGRBs would be produced in delayed mergers between the naked core and the compact object, due to interaction of the binary with a circumbinary disk similarly to the scenario discussed for LFBOTs in section \ref{subsubsec:FBOTs}, or as a result of induced gravitational collapse of the NS to a BH due to accretion of the ejecta mass after the core explodes in a stripped-envelope SN event (e.g., \citealt{RuedaRuffini2012}; \citealt{Fryeretal2014GRBs}). Since the envelope is ejected prior to the explosion(s), such LGRBs will not show hydrogen features in their lightcurves, and their exclusive association with SN Ic-BL (e.g., \citealt{Droutetal2011ICBLSN}; \citealt{Modjazetal2016ICBLSN}; \citealt{Japeljetal2018ICBLSN}; \citealt{Finneranetal2024ICBLSN}) might point towards mergers that occur past helium depletion in the core (e.g., \citealt{Dessartetal2017GRBs}).

A post-CE merger of an NS with the helium core of a giant star has been suggested as a potential explanation for the unusual LGRB 101225A \citep{Thoneetal2011GRB101225A}.This LGRB was relatively low luminosity with a very long gamma-ray emission timescale compared to canonical LGRBs, followed by bright X-ray afterglow with a UV, optical and IR counterparts that appeared as a cooling expanding black body followed by a faint SN-like emission. As in the framework of \cite{BarkovKomissarov2011hypernova} (see section \ref{subsubsec:SNe}), the LGRB could result from a magnetically-driven explosion, whose long timescale could account for the long duration of LGRB 101225A.  The different emission bands could be explained by interaction of the jets that are launched during the merger with different regions of the previously ejected common envelope gas, as it breaks out of it. The small amount of $^{56}\text{Ni}$ synthesized during the merger due to the relatively low temperatures of the accretion disk formed during a merger of a compact object with a helium core could power the faint associated SN, that is best fit with the template of the canonical Type Ic-BL SN 1998bw \citep{Galamaetal1998SN1998bw}. We note that while the possibility of Type Ic SNe to originate in explosions of helium stars that lose their helium envelopes through stellar winds is explored in the literature (e.g., \citealt{Dessartetal2020GRBs}), it is uncertain whether this scenario could be compatible with the energetics and emission lines in Type Ic-BL SNe, posing a challenge for the interpretation of LGRB 101225A as a merger scenario.  


The wide variety of binary properties that lead to these mergers (e.g., \citealt{Sokeretal2019FBOTs}; \citealt{Grichener2023popSynth}) could potentially explain its possible association with different transients that exhibit diverse signatures across the electromagnetic spectrum.

\subsection{High energy neutrino messengers}
\label{subsec:neutrinos}
 
Due to the large opacity of the shared envelope, obtaining direct optical observations of NS/BH-core mergers that occur during the CEE phase is not possible. While modelling the expected ligthcurves (section \ref{subsubsec:GeneralObs}) of these events and exploring potential precursors is crucial for identifying astrophysical transients associated with NS/BH-core mergers, it is insufficient to definitively confirm a transient originated from such a merger. This emphasizes the importance of addressing this challenge with multi-messenger predictions. 

\begin{figure}
\begin{center}
\includegraphics[width=1.0\textwidth]{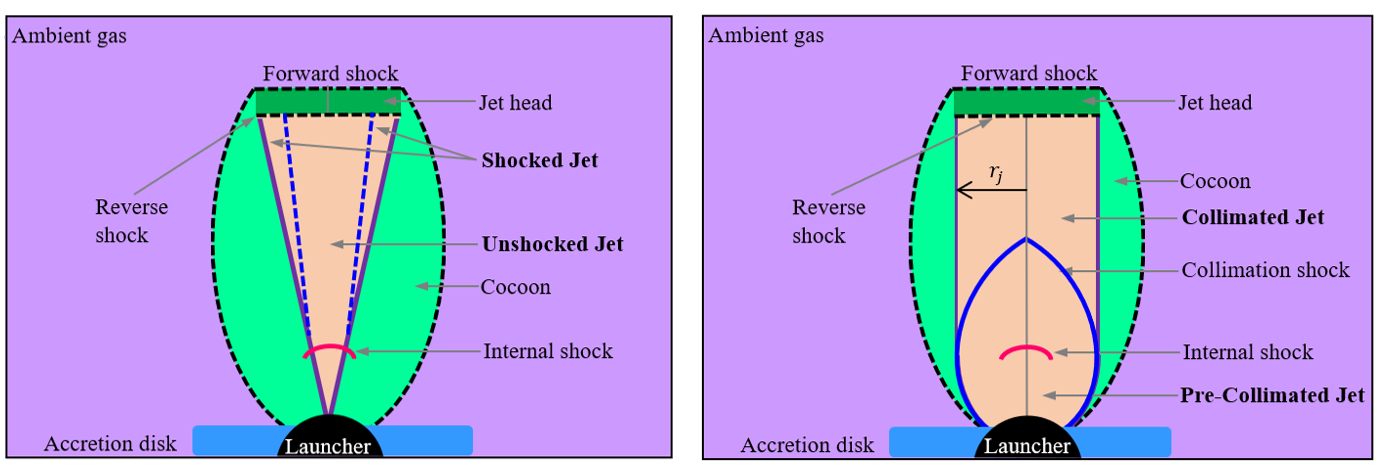}
\caption{A schematic illustration of the geometry of a jet while inside the envelope of a giant star shown on one side of the equatorial plane, adapted from \cite{Brombergetal2011Jets} and \cite{GrichenerSoker2021neutrinos}. Left panel: a conical un-collimated jet that is launched from the BH as it accretes mass from the surrounding envelope. Right panel: the jet is immediately collimated by shocks that are formed as it propagates through the envelope and becomes cylindrical.
}
\label{fig:JetsGeometry}
\end{center}
\end{figure}

\cite{GrichenerSoker2021neutrinos} suggested that high energy neutrinos might be produced through photohadronic interactions that occur in relativistic jets a BH launches prior to its merger with a core of a giant star. When a BH is engulfed by a companion (stage 6 of channel I; left panel of Fig. \ref{fig:EvolutionRoutes}), it begins to spiral-in inside its envelope accreting mass through an accretion disk throughout the in-spiral. The BH launches conical relativistic jets (left panel of Fig. \ref{fig:JetsGeometry}) that deposit their kinetic energy in the envelope gas leading to its expansion. The strong interaction of the jets with the envelope forms a forward shock that propagates into the envelope and a reverse shock facing the supersonic jet. The contact discontinuity between these two shocked regions is termed the \textit{jet head} (dark green area in Fig. \ref{fig:JetsGeometry}). High pressure material in the jet head expands sideways, forming a cocoon around it (light green area in Fig. \ref{fig:JetsGeometry}) that can generate collimation shocks by exerting side pressure (e.g., \citealt{Brombergetal2011Jets}; \citealt{MuraseIoka2013neutrinos}; \citealt{Sennoetal2016neutrinos}). These shocks instantly collimate the jet to a cylindrical shape (right panel of Fig. \ref{fig:JetsGeometry}). Internal shocks could be formed inside the jet as well if fast moving shells manages to catch up with slower shells and merge into them.

The propagation of the collimation and internal shocks in the jet can accelerate protons to very high energies through the diffusive shock acceleration mechanism, also known as first-order Fermi acceleration (e.g., \citealt{Bell1978acceleration}; \citealt{BlandfordOstriker1978acceleration}). The protons repeatedly cross the front of the shock due to scattering caused by magnetic field irregularities. Since the typical timescale for this scattering is much shorter than the timescales for particle collisions in the shock, i.e, the shocks can be regarded as collisionless, the protons can gain energy at the expense of the difference in the kinetic energy between the two sides of the shock. Moreover, \cite{GrichenerSoker2021neutrinos} showed that for the typical luminosities of jets launched inside the CE, the unshocked medium ahead of the shock, known as the upstream of the shock, is optically thin to electron scattering and to pair production, preventing deceleration of the accelerated protons by interactions with lower energy electrons that can be produced through these processes \footnote{\cite{Guarinietal2023neutrinos} study particle acceleration in magnetically driven jets and concludes that if these jets are halted in an extended envelope no efficient particle acceleration can occur at the collimation shocks due to high optical depths, but they do not rule out particle acceleration in internal shocks formed in an hydrodynamical jets choked in an extended envelope. }. Jets produced during the merger itself, however, are much more luminous, and therefore their shocks would be mediated by radiation, preventing efficient acceleration of protons and emission of very energetic neutrinos. 

The accelerated protons usually cool down via photohadronic interactions with thermal photons that are formed in the jet head and diffuse downwards, producing extremely energetic pions mostly through the reaction
\begin{equation}
p + \gamma \rightarrow p + \pi^{0} + \pi^{+} + \pi^{-}.  
\label{eq:Pgamma}
\end{equation}
The newly formed pions are very unstable, and the charged pions decay to muons roughly after $2.6 \times 10^{-8} \sec $ 
\begin{equation}
\begin{split}
\pi^{+} \rightarrow \mu^{+} + \nu_{\rm \mu}
\qquad \text{and} \qquad
\pi^{-} \rightarrow \mu^{-} + \bar{\nu}_{\rm \mu}. \\
\label{eq:PionDecay}
\end{split}
\end{equation}
The mean lifetime of the muons is extremely short as well, leading to a further decay to electrons after $2.2 \times 10^{-6} \sec$
\begin{equation}
\begin{split}
\mu^{+} \rightarrow \bar{\nu}_{\rm \mu} + e^{+} + \nu_{\rm e}
\qquad \text{and} \qquad
\mu^{-} \rightarrow \nu_{\rm \mu} + e^{-} + \bar{\nu}_{\rm e}.
\label{eq:MuonDecay}
\end{split}
\end{equation}
The decay chains are accompanied by the emission of high energy neutrinos and anti-neutrinos whose flavors are determined according to conservation of leptonic number. These neutrinos can then oscillate while steaming out of the envelope (e.g., \citealt{CarpioMurase2020oscillations}). \cite{GrichenerSoker2021neutrinos} find the maximum energy of neutrinos emitted prior to the merger of a BH with the core of the giant to be
\begin{equation}
E_{\rm max,\nu} \simeq 
 2\times 10^{15} \eV  \left( \frac{B'}{10^{3} \G}\right) \left( \frac{T'_{\rm \gamma}}{0.35 \keV}\right)^{-3} \left( \frac{\Gamma}{100}\right),
\label{eq:Enu}
\end{equation}
where $B'$ and $T'_{\rm \gamma}$ are the jet's magnetic field and temperature of the thermal photons that originate in the jet's head, both measured in the jet's rest frame, respectively, and  $\Gamma$ is the Lorentz factor of the jet. Neutrinos of these energies could be detected by \textit{IceCube Neutrino Observatory} (e.g., \citealt{Aartsenetal2013neutrinos}; \citealt{Aartsenetal2021neutrinos}). \footnote{So far only a few objects were confidently linked to high energy neutrinos of these magnitudes: blazar TXS 0506+056 \citep{Ansoldietal2018blazars}, tidal disruption event (TDE) AT2019dsg \citep{Steinetlal2021TDEs}; TDE candidate AT2019fdr \citep{Reuschetal2022TDEs}; TDE candidate AT2019aalc \citep{VanVelzenetal2024TDE}.  }

The neutral pions that are formed through reaction (\ref{eq:Pgamma}) decay producing gamma-rays after $8.4 \times 10^{-17} \sec $
\begin{equation}
\pi^{0} \rightarrow \gamma + \gamma.
\label{eq:NeutralPion}
\end{equation}
While the energy flux of high energy neutrinos and gamma-rays should be comparable, the isotropic diffused gamma-rays emission measured by Fermi gamma-ray telescope \citep{Ackermannetal2015FermiFlux} is much lower than the isotropic neutrino flux inferred by \textit{IceCube} (e.g., \citealt{Aartsenetal2014IceCube}), pointing towards an optically thick high energy neutrino source. \cite{GrichenerSoker2021neutrinos} showed that the jets launched by a BH while spiraling inside the envelope of a giant are choked inside the extended envelope, which is optically thick to gamma-rays in the sub-TeV range, in which the gamma-rays that accompany high energy neutrino emission are expected to be detected due to cooling by interaction with matter (e.g., \citealt{Cellietal2017gammarays}; \citealt{Capanemaetal2021gammarays}).   

According to \cite{Aartsenetal2021neutrinos}, the event rate of high energy neutrino sources with the effective neutrino luminosities found in \cite{GrichenerSoker2021neutrinos} required to explain the entire neutrino flux inferred by \textit{IceCube} is about $3\%$ from CCSNe. \cite{Grichener2023popSynth} performs a population synthesis study and find the event rate of massive binaries where a BH goes into CEE with a giant star to be about $0.6\%$ from CCSNe, indicating that BH-core mergers might have a significant contribution to this flux. Further research is needed to determine the differential and diffused neutrino spectrum expected on earth due to these events. 

The production of high energy neutrinos during mass accretion by a stellar mass BH that resides in a binary system has been studied in the literature in other scenarios as well. \cite{Sridharetal2024Hypernebulae}, for instance, study the emission of high energy neutrinos from winds and jets launched by a compact object that accretes mass from an evolved post-MS companion at a super-Eddington rate following RLOF but shortly prior to CEE. Such accretion can power an hypernebula \citep{SridharMetzger2022hypernebula}, which is an energetic bubble ($\lesssim 5\times10^{7} R_{\rm \odot}$) inflated into the circumbinary medium and is thought to be correlated with recurring fast radio bursts (e.g., \citealt{Sridharetal2021FRBs}). The neutrinos are mostly produced through photohadronic interactions of protons that are accelerated at the termination shocks with photons from the accretion disk. They find that hypernebulae could explain up to $\simeq 25 \%$ of the high energy diffuse neutrino flux inferred by \textit{IceCube}. 

\cite{Guarinietal2022neutrinos} study high energy neutrino emission during the merger of a BH with the stripped core of a giant after a stable RLOF, in a scenario that could result in LFBOTs (\citealt{Metzger2022FBOTs}; see section \ref{subsubsec:FBOTs}). The high mass accretion rates lead to the formation of an accretion disk that ejects outflows in the polar regions. These outflows propagate in the CSM generating shocks where protons can be accelerated. The protons then interact with steady target protons of the CSM, producing high energy neutrinos as they cool. The neutrino flux emitted in this scenario is expected to be below the sensitivity of \textit{IceCube}. Future detection of neutrinos correlated with LFBOTs would be of crucial importance both for probing the origin of high energy neutrinos, and shedding light on progenitors of LFBOTs and perhaps on the physics of BH-core mergers.  

\subsection{Gravitational wave messengers}
\label{subsec:GWs}

Another multi-messenger signature that could be invaluable for associating NS/BH-core mergers with transients is GWs. Over the past years, several studies have been done to explore GW emission from the in-spiral during the CEE phase and its detectabllity prospects with current and near future facilitates. \cite{Renzoetal2021GWs} perform a model free study that focuses on GW emission during the self-regulated phase of CEE and find it is potentially possible to expect at least a few GW detections from Galactic CEE within the \textit{LISA} mission (e.g., \citealt{Thorpeetal2019LISA}). \cite{MoranFraileetal2023GWs} compute the GW signals from the dynamical CEE phase of low mass stars using three dimensional magneto-hydrodynamics simulations and find that the final stages of CEE leading to the merger of the stellar cores are in the frequency band of \textit{LISA} observations. 

\cite{Holgadoetal2018GWs} and \cite{Ginatetal2020GWs} study GW signatures from a CEE phase of an NS with a massive star. While \cite{Holgadoetal2018GWs} focus on the in-spiral and conclude that it might result in GW emission detectable by advanced LIGO as far away as the Magellanic Clouds, with stronger signals than low mass X-ray binaries, \cite{Ginatetal2020GWs} explicitly study GW emission during the merger of the NS with the core. They find that the in-spiral leading to such merger events produces a GW signature that differs significantly from the in-spiral leading to binary compact objects merger. This signature could be observable by space-based GW detectors, but due to the short duration of the event they expect it is likely to be detected with \textit{BBO} (e.g., \citealt{CutlerHolz2009BBO}) rather than \textit{LISA}. We note, however, that since \citealt{Ginatetal2020GWs} do not simulate the feedback of the in-spiral on the surrounding gas, their results might differ from the expected observations. 

Another source of potentially detectable GWs produced during the merger of a compact object with the core of a giant star are Rossby instabilities that form in the accretion disk as a result of sharp density gradients. Non-axisymmetric modes of these instabilities could induce density waves, creating a quadruple moment that generates GWs. \cite{Gottliebetal2024GWs} explore the GW emission from collpsars disks and find they may be detectable in \textit{LIGO-Virgo-Kagra} (e.g., \citealt{Abbottetal2018LVK}; \citealt{Abbottetal2020LVK}) at a rate of about one event per year, implying that existing data might already contain such GW events. Turbulent motions in the jet-cocoon outflow could also result in stochastic GW bursts with a wide range of frequencies. \cite{Gottliebetsl2023GWsfromJets} study the GW bursts powered by cocoons in collapsars and find they emit in the \textit{LIGO-Virgo-Kagra} band and could be detected with third generation facilities like \textit{Cosmic Explorer} (e.g., \citealt{Evansetal2023CosmicExplorer}) and the \textit{Einstein Telescope} (e.g., \citealt{Punturoetal2020EinsteinTelescope}). 

Understanding the detectability prospects of these non-spiralling GW sources in the case of NS/BH-core mergers requires to combine an estimate for their event rates (e.g., \citealt{Schroderetal2020mergers}; \citealt{Grichener2023popSynth}) with the expected frequencies and amplitudes of the GW signal. While NS/BH-core mergers are expected to occur more frequently than collpsars, it is challenging to predict the difference in their GW signature from analytical considerations. Computing the GW patterns that originate in the accretion disk and cocoons formed around the jets during NS/BH-core merger events requires a treatment in the framework of full three dimensional hydrodynamical simulations of these systems due to the asymmetric nature of jet launching. Identifying GW signals that are compatible with predictions of these simulations could be of crucial importance in shedding light on the physics of NS/BH-core mergers and on the jet launching mechanism, and advance our understanding of these transients.

\section{How do we move forward?} 
\label{sec:future}

As discussed in previous sections, since the jets launched inside the core of a giant star as it mergers with the compact object are extremely energetic, these mergers might account for various high energy astrophysical phenomena. Their rate, which is found to be in the range of one event per $100-300$ CCSNe in the case of NS-core mergers and one event per $300-1000$ CCSNe in the case of BH-core mergers for 'fiducial' model parameters (\citealt{Schroderetal2020mergers}; \citealt{Grichener2023popSynth}) combined with their estimated lightcurve properties implies a high probability to detect them with current and upcoming surveys. Since \textit{ZTF} \citep{Bellmetal2019ZTF} and \textit{ASAS-SN} \citep{Kochaneketal2017ASASSN} have been discovering hundreds of SN-like transients yearly, it is possible  that several mergers are already present in their archival data. The prospects for future detection with \textit{LSST} \citep{Ivezicetal2019LSST} are promising, with the potential to observe hundreds of merger events annually, though rapidly evolving transients might be missed. X-ray follow-ups performed with \textit{XMM-Newton} \citep{Jansenetal2001XMMnweton} or \textit{SWIFT} \citep{Gehrelsetal2004SWIFT} could be useful for obtaining information on early CSM interactions, and IR JWST follow-ups could probe dust formation due to the expelled mass during the merger. Our greatest challenge lies in reliably associating observed transients to a merger origin, rather than managing the detection itself.

Modeling NS/BH-core mergers is crucial for advancing towards these associations. Over the years, many studies performed one dimensional and two dimensional simulation mimicking the CEE phase involving a compact object and the core of a giant.(e.g., \citealt{Taametal1978CE1D}; \citealt{Delgadoetal1980CE1D}; \citealt{BodenheimerTaam1984CEE2D}; \citealt{ArmitageLivio2000CEE2D}; \citealt{Gilkisetal2019CEJSNimpostors}; \citealt{Gricheneretal2021NJF}), or the merger itself (e.g., \citealt{Sokeretal2019FBOTs}; \citealt{GrichenerSoker2019rprocess}; \citealt{GrichenerSoker2023W49B}). Performing a wider parameter space exploration with one dimensional simulations could give insight into their typical timescales, nucleosynthesis and locations within their host galaxies. Two dimensional simulations could also be used to study different aspects of the accretion flow and disk formation, as done in previous works (e.g., \citealt{Shimaetal1985twoD}; \citealt{Fryeretal1996MergerDrivenExplosions}; \citealt{BlondinPope2009}; \citealt{Blondinetal2013TwoD}; \citealt{Tripathietal2024TwoD}). Due to the highly asymmetrical nature of jets physics, however, conducting three dimensional hydrodynamical simulations of NS/BH-core mergers is essential for accurately predicting their expected lightcurves and remnant morphologies, which are crucial for confidently linking them to detected transients. 

The subject of mass accretion by a compact object as it spirals trough envelope during the CEE phase has been broadly studied using three dimensional simulations (see \citealt{RopkeDeMarco2023CEEreview} for a recent review). Though several groups studied the evolutionary stages prior to the merger (e.g., \citealt{Schroderetal2020mergers}) or up the formation of an accretion disk (e.g., \citealt{Eversonetal2024TZOs}), to date, there are no three dimensional hydrodynamical simulations that capture jet launching and evolution during NS/BH-core mergers. Performing hydrodynamical simulations of such systems can be very challenging computationally. One of the most problematic numerical difficulties arises from the fact that it contains two processes that occur on very different scales. The Keplerian orbital motion of the compact object is extremely slow compared to the velocity of the sub-relativistic or relativistic jets that it launches. This creates a problem in defining a typical timestep suitable for running the simulation, since setting small enough timesteps as required to resolve the launching and interactions of mildly relativistic jets prevents us from simulating even a considerable fraction of one orbit of the compact object. \cite{LopezCamaraetal2020jets}, for instance, inject the jets that an NS launches in the envelope of a giant star with a velocity of $c/3$ and are able to simulate only a small fraction of the NS's orbit. Even without accounting for the orbital motion, resolving the launching of jets by a compact object can be computationally challenging due to the small scales involved (see section 3 in \citealt{Soker2024CCSN} for a discussion). One way to address this challenge is by using 'subgrid physics' to mimic the effect of jet launching in the simulations, i.e.,  by approximating the collision of the jets' material with its immediate surroundings (e.g., \citealt{Schreieretal2021CEJSN}; \citealt{Hilleletal2022CEJSN}; \citealt{Schreieretal2023CEJSN}; \citealt{Hilleletal2023CEE}). Other complications arise from the variety of physical ingredients that are important for simulating such mergers, such as the self-gravity of the giant star, the magnetic fields prior to the jet launching and in the jets, relativistic effects related to the jet propagation in the envelope or core media, neutrino transport and the effects of radiation pressure and neutrino cooling on the accretion rate by the compact object (see section \ref{sec:physics} for more details). Advances in computational power may enable the execution of such simulations in the future, but even partial simulations that omit some physical processes can still offer invaluable insights for their detection.

Performing three dimensional hydrodynamical simulations of accreting compact objects inside the envelope and core of a giant star will advance research in this field and open new research directions that can lead to a better understanding of these systems and their respective outcomes. Post-processing of these  simulations is the next crucial step to explore the possible processes that can occur inside the accretion disk and the jets it launches. To find the contribution of NS-core mergers to r-process abundances in the Galaxy and Universe (section \ref{sec:rprocess}) we can use the properties of the accretion disk and jet material obtained in these simulations, with emphasis on their neutron richness, and insert them to a nuclear reaction network like \textsc{skynet} \citep{LippunerRoberts2017skynet} to compute the r-process nucleosynthesis yields in such events; The velocity profiles and nucleosynthesis yields from the merger could be given as an input to a radiation hydrodynamic code to calculate the expected lightcurves (section \ref{subsec:EMsignals}); Computing the estimated flux of high energy neutrinos (section \ref{subsec:neutrinos}) emitted prior to a BH-core merger event and giving detectability prospects with \textit{IceCube} (e.g., \citealt{Aartsenetal2013neutrinos}) or \textit{IceCube-GEN2} (e.g., \citealt{Clarketal2021IceCubeGEN2}) could be done by obtaining the energies of the protons accelerated across the shocks interface and taking into account all their possible cooling processes as well as the detailed micro-physics of pion production and decay; The GW signatures and the potential to identify them in the O4 data or with future runs/next generation ground or space-based detectors could be obtained from using retarded Green's functions to post-process the simulations output (section \ref{subsec:GWs}; see the Appendix in \citealt{Gottliebetsl2023GWsfromJets} for an implementation). A simultaneous detection of multiple messengers from the same transient, aligning with predictions from hydrodynamic simulations, could serve as a 'smoking gun' to confirm its origin as an NS/BH-core merger.

\section{Summary}
\label{sec:Summary}

\begin{figure}
\begin{center}
\vspace*{0.5cm}
\includegraphics[width=0.6\textwidth]{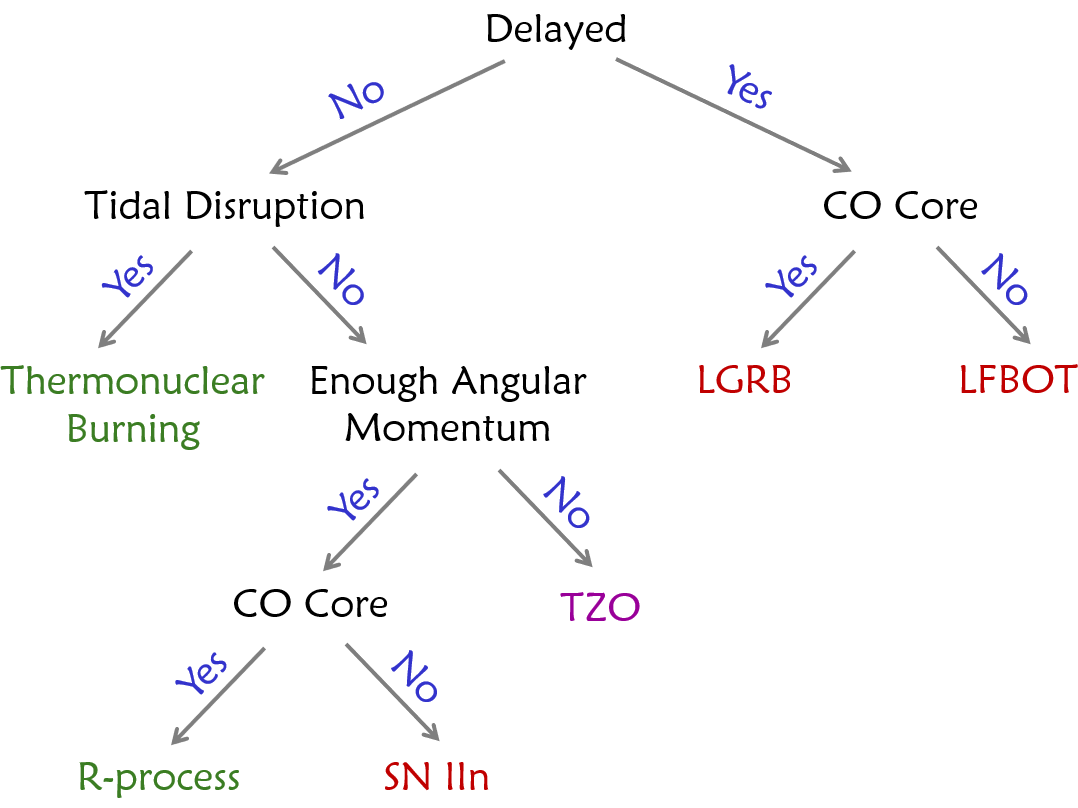}
\vspace*{0.7cm}
\caption{A schematic illustration summarizing the various possible outcomes of NS/BH-core mergers across different parameter regimes, as discussed in this review. The red, purple and green colors represent known observable transients, hypothetical hybrids and outbursts powered by nucleosynthesis, respectively
}
\label{fig:MergersZoo}
\end{center}
\end{figure}

In this review we provided an overview of mergers between compact objects and cores of giant stars, exploring the key findings and developments in this field, and outlining prospects for future advancement. We summarize the key points in Fig. \ref{fig:MergersZoo} and below. 

\textit{Evolutionary pathways}. Massive binaries can evolve into NS/BH-core mergers if the SN explosion in which the primary star produces a compact object does not disrupt the binary system. In such cases the secondary star may engulf the compact object, initiating a CEE phase. Depending on the properties of the compact object-giant binary system at the onset of CEE, it could lead to either the ejection of the shared envelope and formation of a binary compact objects system, or the merger of the compact object with the core of the giant. The merger could occur either through tidal disruption of the core's matter by the compact object before it manages to reach the core's surface, or through disruption of the core by the vast accretion that occurs when the compact object spirals inside it. In both cases the accretion disk that is formed around the compact object could launch jets that power a luminous transient. 

\textit{R-process nucleosynthesis}. If the rate at which an NS accretes mass as it merges with the core of a giant star is high enough, the hot and dense matter of the neutrino-cooled accretion disk that is formed around the NS can self neutronize and produce lanthanides through captures of neutrons on isotopes that originate in cooler regions of the disk. This r-process nucleosynthesis occurs on timescales that are dictated by the evolution of the secondary star ($\simeq$ tens Myrs), potentially allowing NS-core mergers to account for heavy element formation in the early Universe and UFD galaxies, where BNS mergers face challenges. Three dimensional hydro-dynamical simulations are required to properly simulate r-process nucleosynthesis in NS-core mergers.     

\textit{Multi-messenger signatures}. The disk winds and jets launched by the accretion disk formed around the compact object during the merger play a key role in shaping the lightcurve of the resultant transient. They lead to the formation of a highly asymmetric CSM, contribute to the variability of the emission and produce dust with strong IR radiation. The mergers of NSs/BHs with cores of giant stars have been proposed as progenitors of transients that at least partially exhibit these behaviours, such as type IIn SNe, iPTF14hls-like transients, LFBOTs and low-luminosity LGRBs. However, since the gas of the giant’s envelope is very opaque, not allowing for direct optical observations of the underlying physical processes, detecting GW and high energy neutrino signatures correlated with these mergers would be of crucial importance for associating a transient with a merger origin. 

While significant challenges remain in modeling NS/BH-core mergers and achieving their detection, the rapid advancements in computational power and observational capabilities provide a promising outlook for future research in this area.

\section*{ACKNOWLEDGMENTS}
\label{sec:acknowledgments}
I thank Noam Soker for invaluable discussions on the physics of these mergers and their respective outcomes over the years. I thank Brian Metzger and Hagai Perets for eye-opening conversations about the accretion process during CEE. I thank Azalee Bostroem and Manisha Shrestha for helpful input on the observational properties of these systems. I thank Mathieu Renzo for insightful suggestions and discussions that improved this manuscript, as well as the anonymous referee for their valuable comments and recommendations that further improved its quality. I acknowledges support from the Steward Observatory Fellowship in Theoretical and Computational Astrophysics, the IAU-Gruber Fellowship, and the CHE Fellowship.  

\clearpage

\bibliography{refs}

\end{document}